\documentclass[twocolumn]{WileyMSP-template}

\begin{document}

\pagestyle{fancy}
\rhead{\includegraphics[width=2.5cm]{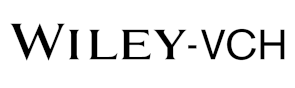}}

\twocolumn[
\begin{@twocolumnfalse}
\title{Non-local correlation and entanglement of ultracold bosons in the two-dimensional Bose-Hubbard lattice at finite temperature}

\maketitle

\medskip

\author{Ulli Pohl, Sayak Ray*, and Johann Kroha}

\medskip

\begin{affiliations}

Physikalisches Institut, Rheinische Friedrich-Wilhelms-Universit\"at Bonn, Nu\ss allee 12, 53115, Bonn, Germany\\
Email: ulli.pohl@uni-bonn.de, sayak@uni-bonn.de, kroha@th.physik.uni-bonn.de

\end{affiliations}


\medskip

\end{@twocolumnfalse}
]

\begin{sloppypar}

\begin{abstract}

We investigate the temperature-dependent behavior emerging in the vicinity of the superfluid (SF) to Mott-insulator (MI) transition of interacting bosons in a two-dimensional optical lattice, described by the Bose-Hubbard model. The equilibrium phase diagram at finite-temperature is computed using the cluster mean-field (CMF) theory including a finite-cluster-size-scaling. The SF, MI, and normal fluid (NF) phases are characterized as well as the transition or crossover temperatures between them are estimated by computing physical quantities such as the superfluid fraction, compressibility and sound velocity using the CMF method. We find that the non-local correlations included in a finite cluster, when extrapolated to infinite size, leads to quantitative agreement of the phase boundaries with quantum Monte Carlo (QMC) results as well as with experiments. Moreover, we show that the von Neumann entanglement entropy within a cluster corresponds to the system's entropy density and that it is enhanced near the SF-MI quantum critical point (QCP) and at the SF- NF boundary. The behavior of the transition lines near this QCP, at and away from the particle-hole (p-h) symmetric point located at the Mott-tip, is also discussed. Our results obtained by using the CMF theory can be tested experimentally using the quantum gas microscopy method.

\end{abstract}

\section{Introduction}

\label{Intro}

The physics of strong correlation and entanglement in interacting quantum systems has been a focus of research in condensed matter and statistical physics for a long time, covering a broad area ranging from equilibrium to out-of-equilibrium phenomena \cite{Sachdev, Tauber17, Lewenstein12, Fisher67, Stanley, Horodecki09}. In recent years, ultracold atomic systems have become an ideal platform to explore this field of research due to the fine-tunability of system parameters, so that various correlated model Hamiltonians exhibiting a quantum phase transition can be realized \cite{Zwerger08}. Recent experimental tools and imaging techniques have enabled to explore the density profiles of ultracold atoms with single-lattice-site resolution \cite{Kuhr10, Greiner09}, as well as have provided access to {\it in situ} measurements of local observables such as density, density fluctuation, correlation functions and entanglement \cite{Chin09, Chin11, Chin12, Greiner15, Greiner19}. These developments have opened an avenue to analyze the strongly correlated phases, phase transitions and non-equilibrium dynamics using cold atom systems. 

The Bose-Hubbard model (BHM) is known for its success to describe the phases and dynamics of ultracold interacting bosons in a lattice \cite{Fisher89, Zoller98}. Its experimental realization in an optical lattice led to the landmark demonstration of the superfluid (SF) to Mott insulator (MI) phase transition \cite{Greiner02} and its subsequent observations \cite{Spielman07, Spielman10}. Over the past few decades, a number of theoretical studies have investigated the different phases of the BHM at zero or the lowest possible temperature by means of different methods \cite{Stoof01, Monien96, Freericks13, Pelster09, Pelster09_1, Rozek15, Dupuis11, Dupuis11_1, Dupuis13, Dupuis13_1, Svistunov07, Svistunov08, Rigol09, Trivedi11, Vollhardt08, BDMFT11, Sengupta12, Monien98, Monien00, Kollath04, Kollath05, Knap05}.
The universal conductivity at the SF-MI phase transition in two dimension has also been analyzed in a quantum Monte Carlo (QMC) study and by using the AdS/CFT correspondence \cite{Sachdev14, Prokofev14}.
The single-particle spectra in the different phases, for instance the particle-hole (p-h) excitation gap in the Mott insulator, as well as collective modes, like the gapless Goldstone mode and the massive Higgs mode in the superfluid phase, have been calculated \cite{Monien99, Stoof03, Sengupta05, Blatter07, Auerbach11}, and their existence have been detected in cold-atom experiments for bosonic and fermionic 
gases \cite{Bloch12, Koehl18}.

The behavior at finite temperatures is far less explored. There, apart from an incompressible MI and a compressible SF phase, a normal fluid (NF) with vanishing superfluidity but finite compressibility appears as a consequence of a transition from the superfluid or a crossover from the Mott insulator, respectively \cite{Fisher89}. 
The SF-NF transition and the behavior of other thermodynamic quantities have been investigated in a few theoretical studies \cite{Trivedi08, Wessel16, Mueller11, Wang11, Duchon13, Trivedi12, Joshi20, Pires17}. Experimentally, the reduction of $T_c$ near the QCP has been observed across a vacuum-to-superfluid transition \cite{Chin11, Chin12} and across the MI-SF transition at a constant particle density (particle number per lattice site) $\bar{n}=1$ \cite{Troyer10}. However, a systematic analysis on the effect of correlation at low temperature is lacking, particularly in 2D systems where simple mean-field and perturbative methods fail due to enhanced fluctuations, as well as exact solution methods like density matrix renormalization group (DMRG) are not easily available. 

In this paper we primarily investigate the effect of non-local 
correlation at low temperature in the two-dimensional Bose-Hubbard 
model by means of a computationally inexpensive approach, namely, the
cluster mean-field (CMF) theory \cite{Demler12, Yamamoto12, Luhmann13, Carlos12, Vezzani05, Morigi20, Kuldeep20, Ray20, Silva19, Fazio16}. 
The finite-size clusters used in the CMF method incorporate spatial
correlation which is beyond the scope of single site mean-field theory, 
as well as the single site bosonic dynamical mean-field theory (B-DMFT) 
where dynamical fluctuations are incorporated locally \cite{Vollhardt08, BDMFT11}. 
We extend the CMF method to finite temperature, and study, particularly, the SF--NF transition, characterized by the superfluid transition temperature $T_c$, and the MI--NF crossover, characterized by the crossover temperature $T^*$ near the QCP, both at and away from the p-h symmetry point of the BHM. While the latter one is the generic transition achieved by keeping the chemical potential fixed, the former one is a special case where the average particle number per lattice site is kept fixed at $\bar{n} = 1$. The equilibrium phase diagram is computed for both the cases. As one might expect from any MF like methods, also CMF method can not capture the behavior in the critical region of the Berezinskii--Kosterlitz--Thouless (BKT) transition relevant for the model at hand \cite{Berezinskii72, KT73, Nelson77}.
Our main results, as an application of the CMF method to the paradigmatic BHM in 2D at finite temperature are summarized below. 
\\
(I) We show that the method, after an appropriate cluster-size scaling, quantitatively reproduces the phase boundaries and also their associated critical exponents in agreement with the QMC results, specifically from Refs.~\cite{Svistunov08, Trivedi11, Trivedi12}. 
\\
(II) We calculate the relevant physical quantities such as superfluid fraction, sound velocity and compressibility using the CMF method. Compressibility provides access to the the non-local density correlations and to the particle and hole excitation gaps in the MI phase which agree quantitatively with the experiment in Ref.~\cite{Bloch12}. 
\\
(III) We further compute the von Neumann entanglement entropy within a cluster, namely, {\it cluster entanglement entropy} (CEE), exhibiting enhancement near the QCP and at the SF-NF boundary. We show that the CEE corresponds to the system's entropy density, and thereby, it provides a robust signature of the transition. 
\\
Our theoretical results are also consistent with the experimental findings in the vicinity of the density-induced vacuum-to-superfluid transition \cite{Chin11, Chin12}.

The paper is organized as follows. In Sec.\ \ref{Model} we introduce the BHM and explain the CMF technique. Next, in Sec.\ \ref{QCP_generic} we present the finite temperature phase diagram and discuss the behavior of $T_c/U$ and $T^*/U$ near the QCP for a generic, non-p-h symmetric MI-SF transition. This is followed by the estimation of the Mott gap and comparison with QMC and experimental results in Sec.\ \ref{exp_QMC}. The behavior of the phase boundaries at the p-h symmetric multicritical point is discussed in Sec.\ \ref{n1_Tc}. Finally, we summarize and conclude in Sec.\ \ref{summary}.

\section{Model and Method}
\label{Model}

\subsection{The Bose-Hubbard model}

The BHM within a single band and tight-binding approximation can be described by the Hamiltonian,
\begin{equation}
\hat{\mathcal{H}} = -\sum_{\langle i,j\rangle} (t\hat{a}_i^{\dagger}\hat{a}_j + \text{h.c.}) - \sum_{i} \left[\mu \hat{n}_i + \frac{U}{2} \hat{n}_i(\hat{n}_i-1) \right]    
\label{Ham}
\end{equation} 
where $\hat{a}_i^{\dagger} (\hat{a}_i)$ are the bosonic creation (annihilation) operators at the $i$th site, $\hat{n}_i$ represents the local number operator, $t$ is the hopping amplitude between the nearest neighbor (NN) sites of a square lattice denoted by $\langle i,j\rangle$, $U$ and $\mu$ are the onsite interaction, and the chemical potential respectively. In this paper we set the Planck constant $\hbar = 1$, the Boltzmann constant $k_B = 1$, and measure all energies in the unit of $U$ unless it is otherwise mentioned.  

To simulate the BHM \eqref{Ham} on a 2D square lattice, we will adopt the CMF method which was first introduced to study correlated spins and bosonic systems at zero temperature primarily to compute the phase boundaries \cite{Demler12, Yamamoto12, Luhmann13, Carlos12, Vezzani05, Morigi20}. More recently, it has been used to investigate the extended BHM in regular and frustrated lattices \cite{Kuldeep20, Ray20}, as well as to study dynamics of interacting spin systems \cite{Silva19, Fazio16}. 
Here we extend the method to study the behavior around the SF-MI QCP of the 2D BHM as a function of temperature.

\subsection{Cluster mean-field theory}
\label{Method}

\begin{figure}[t]
\centering
\includegraphics[width=\columnwidth]{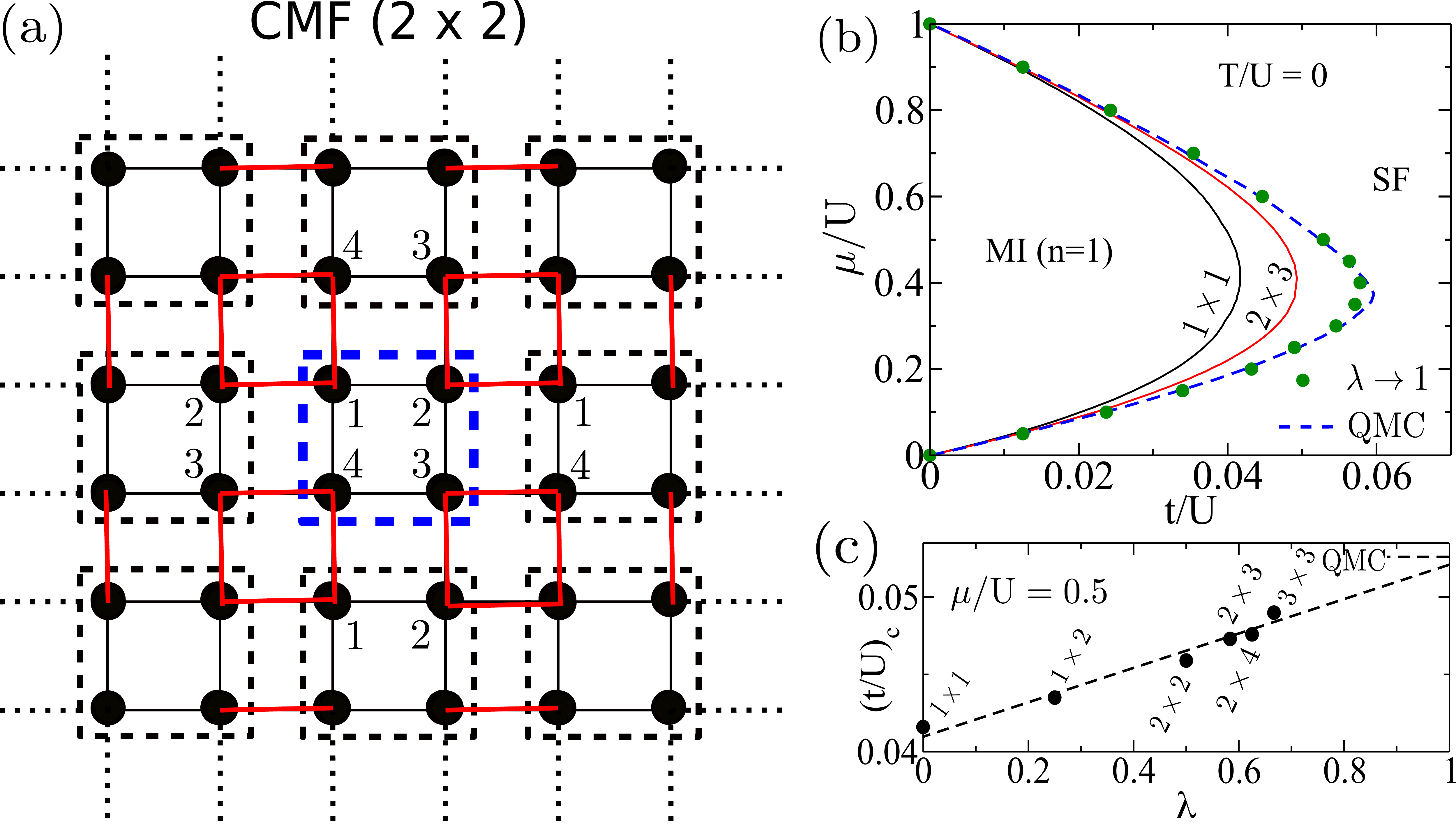}
\caption{(color online) (a) Schematic demonstration of the CMF method in a square lattice split up into $2\times 2$ clusters (boxes with dashed lines). The red lines show the bonds between neighboring clusters. See the text for details of the construction. (b) Zero-temperature phase diagram of the 2D BHM obtained using different cluster sizes as indicated. The critical $(t/U)_c$ in the thermodynamic limit (filled circles), extracted from the cluster finite-size scaling, are compared with QMC results of Ref.\ \cite{Svistunov08} (dashed line). An an example of such extrapolation for $\mu/U=0.5$ is shown in (c).}
\label{CMF_schematic}
\end{figure}

In CMF theory, the entire lattice is decomposed into clusters $\mathcal{C}_1, \mathcal{C}_2, \cdots$ as depicted in Fig.\ \ref{CMF_schematic} (a) for the example of $2 \times 2$ clusters in a 2D lattice.
For a cluster $\mathcal{C}_l$ the field operators on neighboring clusters
$\hat{a}_j$, $j\notin \mathcal{C}_l$, are approximated by their thermal
averages, i.e., the local condensate amplitude, computed with the CMF density matrix
to be determined below.
This leads to the CMF Hamiltonian $\hat{\mathcal{H}} = \sum_l \hat{\mathcal{H}}_{\mathcal{C}_l}$ which is the sum of cluster Hamiltonians,
\begin{eqnarray}
\hat{\mathcal{H}}_{\mathcal{C}_l} &=& -t\bigg[\sum_{\substack{\langle i,j\rangle \\ i,j \in \mathcal{C}_l}} \hat{a}_i^{\dagger}\hat{a}_j + \sum_{\substack{\langle i,j\rangle \\ i \in \mathcal{C}_l, j \notin \mathcal{C}_{l}}} \hat{a}_i^{\dagger}\langle\hat{a}_j\rangle \bigg] + \text{h.c.} \nonumber \\
&-& \sum_{i \in \mathcal{C}_l} \left[\mu \hat{n}_i + \frac{U}{2} \hat{n}_i(\hat{n}_i-1) \right]\ .
\label{ham_cluster}
\end{eqnarray}
Accordingly, the total CMF density matrix factorizes,
\begin{equation}
  \hat{\rho} = \prod_l \hat{\rho}_{\mathcal{C}_l}, \quad \hat{\rho}_{\mathcal{C}_l} = e^{-\beta \hat{\mathcal{H}}_{\mathcal{C}_l}}/\mathcal{Z_{C}}\ ,
\label{density_matrix}
\end{equation}
where $\hat{\rho}_{\mathcal{C}_l}$ is the thermal density matrix of cluster
${\mathcal{C}_l}$ at the inverse temperature $\beta=1/T$, and
$\mathcal{Z_{C}}=\textrm{Tr}\hat{\rho}_{\mathcal{C}_l}$ is the corresponding
partition function of the cluster. Because of translation symmetry, the local
condensate amplitude on neighboring cluster sites, $\langle a_{j}\rangle$,
$j\notin C_l$, are identical to those on sites inside the cluster $C_l$,
shifted by one cluster length as indicated by the site numbers in Fig.~\ref{CMF_schematic} (a). The cluster Hamiltonian in Eq. (\ref{ham_cluster}) is, thus, solved by exact diagonalization and self-consistently computing the thermal averages
$\langle a_j\rangle =\textrm{Tr}(\hat{a}_j\hat{\rho})$ in
Eqs. (\ref{ham_cluster}) using Eqs. (\ref{ham_cluster}) and
(\ref{density_matrix}). After self-consistent diagonalization, any CMF thermal
expectation value can be calculated as $\langle\, .\,\rangle={\rm
  Tr}(\,.\,\hat{\rho})$. One can also compute the total free energy for a
cluster from $F = -T\ln \mathcal{Z_C}$ and derive thermodynamic expectation
values from it.

The relevant quantities which characterize the superfluid at zero temperature are the finite condensate amplitude $\alpha_{\rm SF}$ and the finite compressibility $\kappa$, defined as follows,
\begin{equation}
\alpha_{\rm SF} = \frac{1}{N_{\mathcal{C}}}\sum_{i \in \mathcal{C}} |\langle \hat{a}_i\rangle|, \quad \kappa = \frac{1}{\bar{n}^2} \frac{\partial \bar{n}}{\partial \mu}\ ,
\end{equation}
where the average particle density is, $\bar{n} = (1/N_{\mathcal{C}}) \sum_{i\in \mathcal{C}} \langle \hat{n}_i\rangle$, and $N_{\mathcal{C}}$ is the number of lattice sites within a cluster $\mathcal{C}$.
However, in two dimensions at finite temperature, due to the logarithmic decay of correlation functions with distance, the condensate amplitude vanishes in the thermodynamic limit \cite{Wagner66, Hohenberg67} and, therefore, cannot be used as an order parameter of the SF phase. A physical quantity which does characterize the superfluid, both at zero and finite temperatures, is the SF density $\rho_s$. CMF theory, like other MF theories, yields a condensate amplitude $\alpha_{\rm SF}$ which remains finite in the superfluid, and vanishes at the same transition line to the
MI or NF phase, where the SF density $\rho_s$ vanishes, see Fig.~\ref{mu_t_PD}.
In the following, for simplicity we will therefore use $\alpha_{\rm SF}$ to characterize the SF-MI/NF boundary at zero/finite temperature.

The zero-temperature phase diagram of the BHM in 2D is shown in
Fig.\ \ref{CMF_schematic} (b). It consists of two distinct phases -- an incompressible Mott
insulator, and a compressible superfluid. 
Noticeably, with increasing cluster size, the MI-SF phase boundary is improved over the single-site MF theory obtained simply by using a $1\times 1$ cluster.
For infinite cluster size (thermodynamic limit), the number of bonds in the cluster
is given by $N_b=N_{\mathcal{C}}z_c/2$, where $z_c$ is the lattice
coordination number ($z_c=4$ for the 2D square lattice). Therefore, we employ
cluster size scaling of data with the parameter $\lambda =
N_b/(N_{\mathcal{C}}z_c/2)$ as introduced in Ref.\ \cite{Yamamoto12}, and
extract the values in the thermodynamic limit by extrapolating to $\lambda
\rightarrow 1$, see Fig.\ \ref{CMF_schematic} (c). The
extrapolated critical hopping $(t/U)_c$ for the MI-SF transition is plotted 
as a function of $\mu/U$ in Fig.\ \ref{CMF_schematic} (b) which agrees 
well with the QMC result. The zero-temperature phase
diagram within CMF method has been analyzed in more detail in Ref.\ \cite{Demler12, Luhmann13}. 
Physical results shown in this article will represent this extrapolation 
to infinite cluster size limit, $\lambda\to 1$, unless indicated 
otherwise.

\textit{Fock space truncation and numerical accuracy.}
Since for bosonic systems the occupation numbers in the canonical ensemble
are unlimited, in representing and diagonalizing the Hamiltonian in 
Eqs.~(\ref{ham_cluster}), (\ref{density_matrix}),
the Hilbert space dimension must be truncated not only by 
cluster size but also by the Fock space dimension. We restrict the local 
occupation number on any site to at most $n_i=2$. 
Since higher occupation numbers are exponentially suppressed 
with $n_i(n_i-1)\,U/T$, this approximation turns out to be sufficient for 
low temperatures $T/U \lesssim 0.1$ near the QCP, see Appendix\ \ref{AppenA} for details. For clusters up to $2\times 3$, the full Hamiltonian matrix is diagonalized exactly. For faster diagonalization and convergence of larger clusters at low temperatures, we use a Lanczos algorithm to compute the $\nu_{\rm max}$ lowest-lying eigenstates. For each cluster and temperature,  
$\nu_{\rm max}$ is determined such that the contribution of the higher-energy eigenstates to the partition function is less than $\mathcal{O}(10^{-8})$,
$\exp[-\beta E_{\nu_{\rm max} }]/\mathcal{\tilde{Z}_C} < 10^{-8}$,
where $\mathcal{\tilde{Z}_C}=\sum_{\nu=1}^{\nu_{\rm max}}\exp[-\beta E_{\nu}]$ is the
truncated partition function. For instance, for $T/U=0.1$ we find that
$\sim 20$ \% of the lowest lying states are sufficient to satisfy this criterion.
The computations for fixed density $\bar{n}=1$ discussed in 
Sec. \ref{ph_symmetry} require an additional self-consistency to adjust the chemical potential accordingly. This can substantially increase the 
computation time, but does in principle not limit the numerical accuracy.

\section{Finite temperature phase diagram away from particle-hole symmetry}
\label{QCP_generic}

At a finite temperature, in addition to the MI and SF another phase appears, the
normal fluid (NF), characterized by a vanishing superfluid density $\rho_S$ and vanishing condensate amplitude $\alpha_{\rm SF}$, but finite compressibility $\kappa$. 
In two dimension the transition from superfluid to normal fluid belongs to the Berezinskii--Kosterlitz--Thouless (BKT) universality class corresponding to a vortex binding-unbinding transition \cite{Berezinskii72, KT73, Nelson77}. The study of BKT physics and its associated scaling behavior discussed in Refs.~\cite{Laloe07, Svistunov01, Shlyapnikov00, Rigol12, Kawashima97}, is beyond the scope of the present study, as it requires significantly larger system sizes of at least order of the vortex radius. Nevertheless, the CMF theory gives an overall correct description except for the critical region where BKT physics sets in. In the subsequent sections, we will show that it describes the phase transition line near the QCP in quantitative agreement with QMC.

\begin{figure}[t]
\centering
\includegraphics[width=\columnwidth]{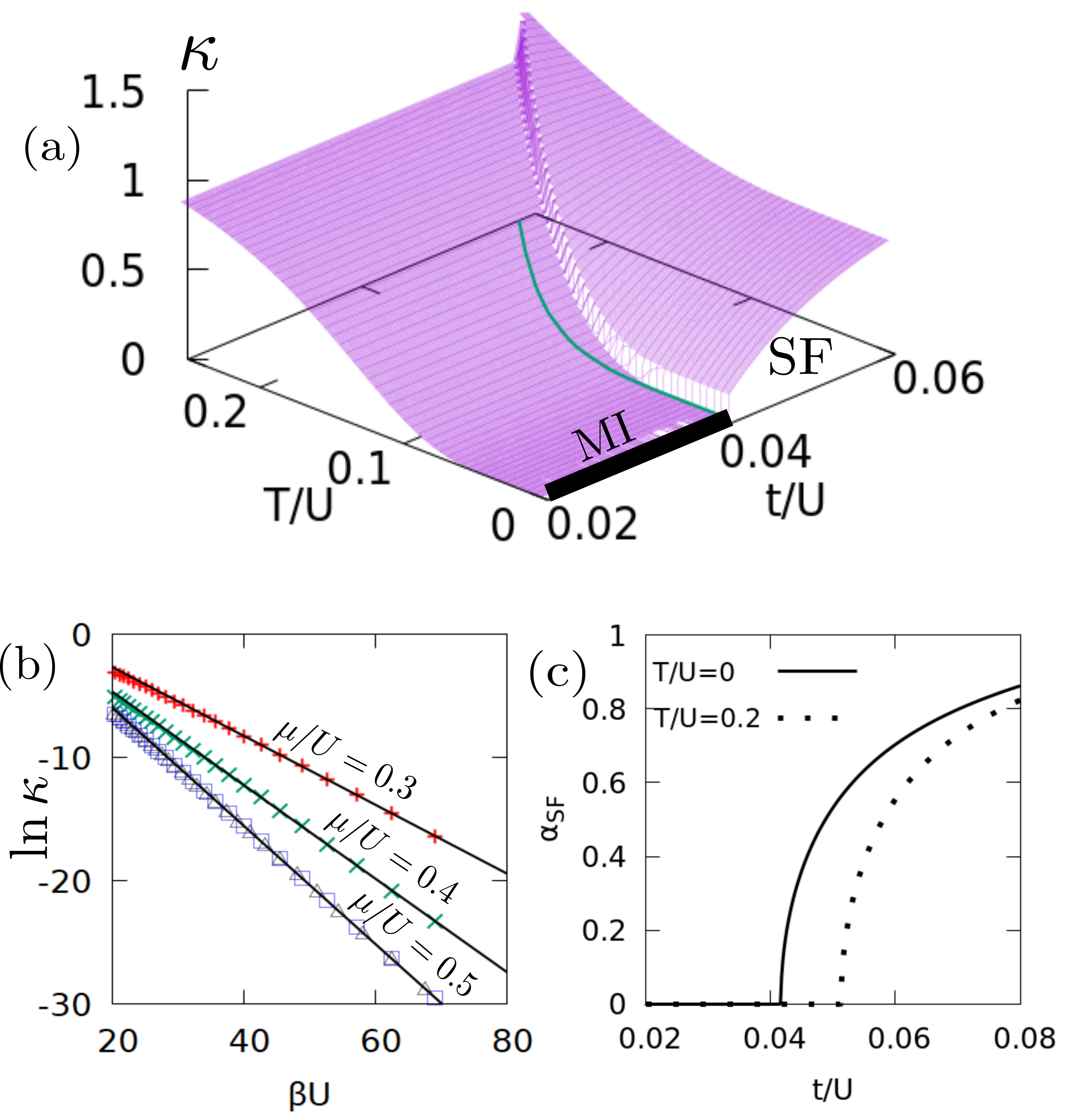}
\caption{(Color online) Single-site MF results. (a) Compressibility $\kappa$
  as a function of $t/U$ and $T/U$ exhibiting a jump at the SF-NF boundary
  where the condensate amplitude $\alpha_{\rm SF}$ vanishes (marked by the
  solid line in the $t/U-T/U$ plane). (b) Semi-log plot of $\kappa$ vs $\beta U$
  at $t/U=0$. The linear fits for each $\mu/U$ are shown by the solid
  lines. In single-site MF, the slope $\Delta_{\rm ph}/U$ does not change with
  $t/U$ as evident from overlapping data of $t/U=0$ (open squares) and
  $t/U=0.01$ (open triangles) at $\mu/U=0.5$. (c) The variation of
  $\alpha_{\rm SF}$ with $t/U$ at sections $T/U=0$ and $T/U=0.2$ is shown. In
  all plots except panel (b) we set a typical value of the chemical potential
  $\mu/U=0.5$, which is away from the p-h symmetry ($t/U \ne 0$).}
\label{FiniteT_MF}
\end{figure}

Another characteristics of the SF and NF phases are the total number fluctuations $\langle \delta \hat{N}^2\rangle$. 
Since the total number operator is $\hat{N} = \sum_i{\hat{n}_i}$, this involves non-local density correlations $\langle \hat{n}_i\hat{n}_j\rangle$. 
In Refs.\ \cite{Wang11, Trivedi12} a relation between $\kappa$ and $\langle \delta \hat{N}^2\rangle$ via the fluctuation-dissipation theorem (FDT) has been studied in order to explore the thermometry of 2D BHM.
While single-site MF theory incorporates only local number fluctuations, $\langle\delta \hat{n}_i^2\rangle =
\langle \hat{n}_i^2\rangle - \langle \hat{n}_i\rangle^2$, the finite cluster mean-field theory does capture the non-local contributions as presented below.

\subsection{Single-site mean-field results}

The MF phase diagram is shown in Fig.\ \ref{FiniteT_MF} (a). 
At zero temperature, the SF-MI transition with vanishing 
compressibility $\kappa$ in the MI phase is visible, which turns into a
SF-NF transition at any finite $T>0$. The SF-NF boundary is depicted by 
the solid line in the $t/U-T/U$ plane in Fig.\ \ref{FiniteT_MF} (a). 
Here and in the following we denote the distance to the QCP by $\delta =
(t/U)-(t/U)_c$. Within MF, at the critical value $(t/U)_c$, the condensate 
amplitude $\alpha_{\rm SF}$ vanishes as $\sim|\delta|^{1/2}$ 
[see Fig.\ \ref{FiniteT_MF} (c)], whereas the response 
quantity compressibility $\kappa$ exhibits a jump [see Fig.\ \ref{FiniteT_MF} 
(a)], indicating a second-order phase transition for any $T\geq 0$. 
At $T=0$, the magnitude of the jump can be analytically estimated by
perturbation theory \cite{Stoof01}. On the MI side, upon
increasing temperature a crossover to the NF occurs,
exhibiting an exponential increase of $\kappa \sim \exp(-\beta \Delta_{\rm
  ph})$ due to thermal activation. Here, $\Delta_{\rm ph}$ is the minimum
energy for adding or removing a particle, that is, 
the particle or hole gap of the Mott insulator, whichever is smaller. 
Fig.\ \ref{FiniteT_MF} (b) shows a logarithmic plot of $\kappa$ as a
function of $\beta U$ for different chemical potentials $\mu/U$. A linear fit
of the data allows to extract $\Delta_{\rm ph}/U$ for each $\mu/U$ in the
limit $\beta U \gg 1$ [see Fig.\ \ref{FiniteT_MF} (b)]. It is not surprising
that the single-site MF theory can give the correct particle or hole gap
only in the atomic limit $t/U=0$, namely, $\Delta_{\rm ph}/U = (1-\mu/U)$ and
$\mu/U$, respectively. At finite $t/U$, the effect of
non-local density fluctuations on $\Delta_{\rm ph}/U$ is not captured. 
Thus $\Delta_{\rm ph}/U$ 
does not change with $t/U$ as illustrated in
Fig.\ \ref{FiniteT_MF} (b) for $t/U=0$ and $0.01$. 
In the next section, we will analyze the variation of
$\Delta_{\rm ph}/U$ with $t/U$ using a finite-cluster calculation, and will
set this p-h excitation energy scale as the crossover temperature between 
the MI and the NF.

\begin{figure}[t]
\centering
\includegraphics[width=\columnwidth]{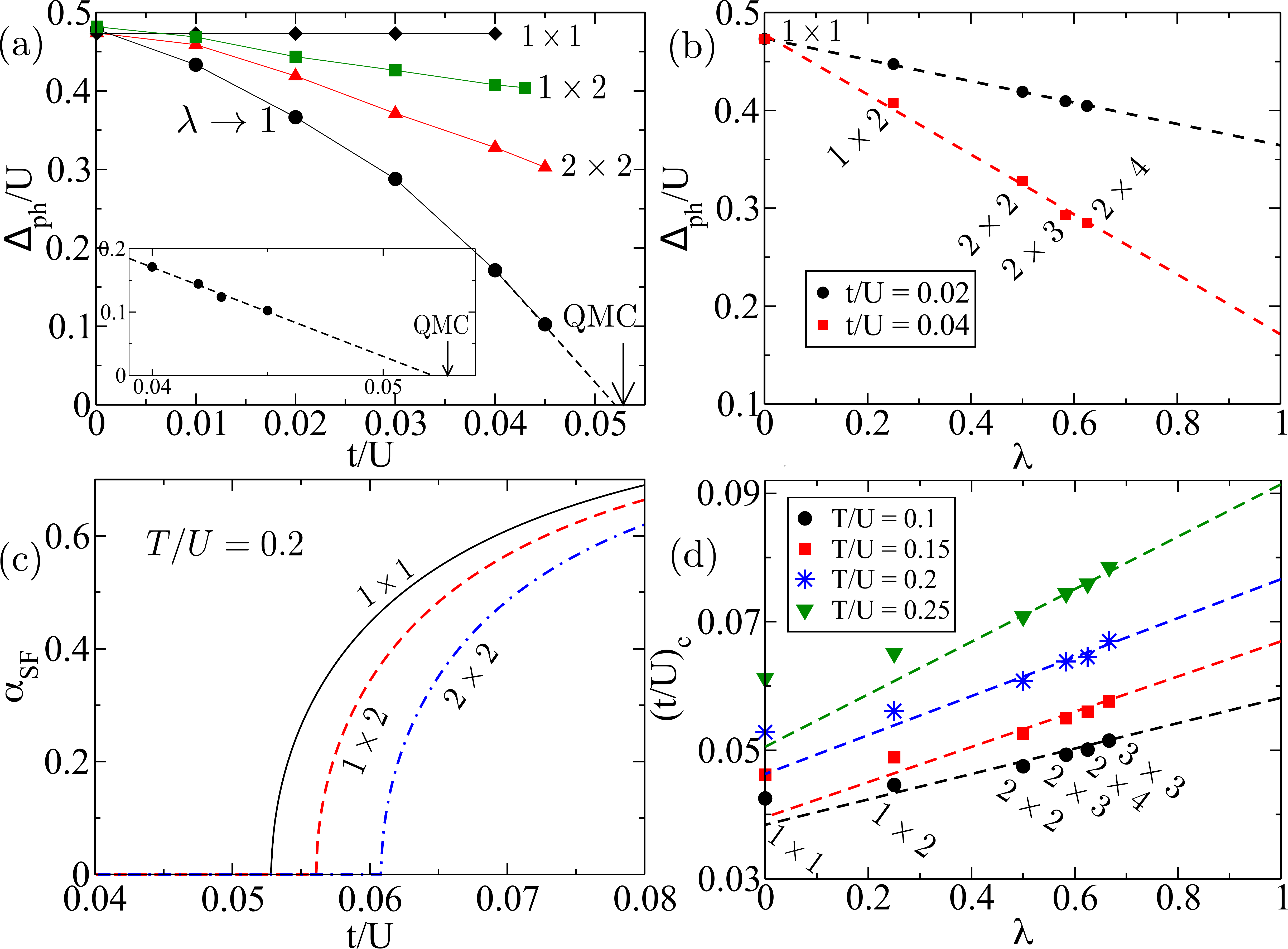}
\caption{(Color online) (a) The particle or hole gap $\Delta_{\rm ph}/U$ 
in the MI at $T=0$, and (c) the condensate amplitude $\alpha_{\rm SF}$ in the SF are
plotted as a function of $t/U$ for increasing cluster sizes. 
The infinite-cluster-size extrapolations of $\Delta_{\rm ph}/U$ 
and of the critical hopping $(t/U)_c$ for SF-NF transition are shown
in (b) and (d), respectively. The inset in (a) shows the linear behavior of
$\Delta_{\rm ph}/U$ extracted at $\lambda \rightarrow 1$ and its linear 
extrapolation to the MI-SF transition. From its linear fitting the critical 
hopping where $\Delta_{\rm  ph}/U$ vanishes is determined as $(t/U)_c \approx
0.0519$ at $T/U = 0$.}
\label{FiniteT_CMF}
\end{figure}
 
\subsection{Non-local correlation effects near QCP}
\label{nonlocal_corr}

As has been shown in Refs.~\cite{Wang11,Trivedi12}, non-local correlations 
contribute to the compressibility via the FDT and are,
thus, expected to influence $\Delta_{\rm ph}/U$ as well.
In a finite-cluster calculation, inter-site correlations such as $\langle
\hat{n}_i\hat{n}_j\rangle$, $i, j \in \mathcal{C}$, are taken into account. 
In Fig.~\ref{FiniteT_CMF} (a) and (c) we show the particle or hole gap
$\Delta_{\rm ph}/U$, whichever is smaller, and the condensate amplitude 
$\alpha_{\rm SF}$, respectively, as a function of $t/U$ for several cluster
sizes. We take the thermodynamic limit by the infinite-cluster extrapolation
$\lambda\to 1$, as illustrated for $\Delta_{\rm ph}/U$ as well as for $(t/U)_c$ (where $\alpha_{\rm SF}$ vanishes)
in Fig.~\ref{FiniteT_CMF} (b) and (d), respectively. It is clearly seen 
that $\Delta_{\rm ph}/U$ depends on $t/U$ even within the MI and NF phases 
where $\alpha_{\rm SF}=0$. As the MI-SF QCP is approached, calculating 
$\Delta_{\rm ph}/U$ would require larger and larger 
clusters as the correlation length diverges. We observe, however, that 
for its smallest values $\Delta_{\rm ph}/U$ depends linearly 
on $t/U$ [see inset of Fig.~\ref{FiniteT_CMF} (a)]. Therefore, we may 
extrapolate $\Delta_{\rm ph}/U$ linearly to zero. The QCP determined in this 
way both from the MI side (vanishing of $\Delta_{\rm ph}/U$) and from the
SF side (vanishing of $\alpha_{\rm SF}$ at $T=0$) in the $\lambda\to 1$ limit is found to
be at $(t/U)_c \approx 0.0519$. Notably, this is in close agreement with 
the QMC result $(t/U)_c \approx 0.0524$, marked by an arrowhead in 
Fig.\ \ref{FiniteT_CMF} (a,d). The linear vanishing of $\Delta_{\rm ph}/U$
can be understood from the behavior 
\begin{equation}
\Delta_{\rm ph}/U\sim |\delta|^{z\nu} \ ,
\label{Delta_critical}
\end{equation} 
with $z\nu=1$ as expected from a generic MI-SF transition away from 
p-h symmetry. Here, $z$ is the dynamical critical exponent, and $\nu$ the 
correlation-length exponent. 

\begin{figure}[t]
\centering
\includegraphics[width=0.85\columnwidth]{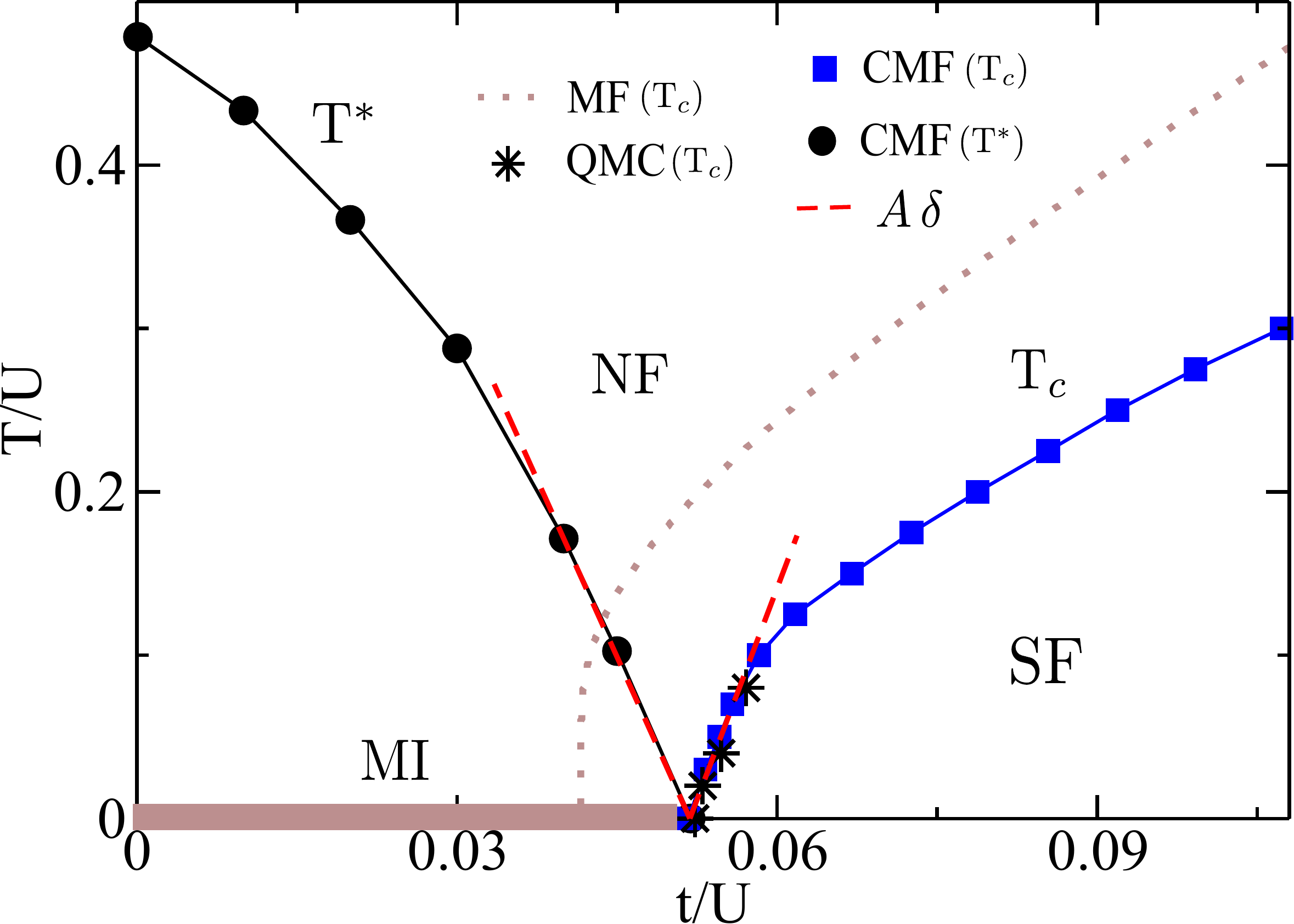}
\caption{(Color online) Generic finite-temperature phase diagram 
  of the 2D BHM for a
  fixed chemical potential $\mu/U=0.5$. The MI phase at $T/U=0$ is marked by
  the bold line on the horizontal axis. The crossover temperature $T^*$ is set
  by the particle or hole gap $\Delta_{\rm ph}$ of the Mott insulator. The
  SF-NF boundary $T_c$ is obtained from the vanishing of $\alpha_{\rm SF}$
  (see also section \ref{sec:SF_fraction}). All
  the data presented here are obtained from infinite cluster size
  extrapolation shown in Fig.\ \ref{FiniteT_CMF}. The red dashed lines are 
  linear fits of the data indicating a linear vanishing of $T^*/U$ 
  and $T_c/U$ near the QCP. QMC data ($*$) from Ref. \cite{Trivedi11} and MF results
  (grey dotted curve) are shown for comparison.}
\label{FiniteT_PD}
\end{figure}

We extend the calculations to finite temperature in order to determine the SF-NF boundary in the thermodynamic limit
from the vanishing of condensate amplitude, followed by the cluster finite-size
scaling, $\lambda\to 1$, in the same spirit as the zero-temperature CMF calculations performed in Ref.\ \cite{Yamamoto12}. The extension of the calculations to finite temperature are shown in Figs.\ \ref{FiniteT_CMF} (c, d).
The linear extrapolation of $(t/U)_c$ for different
clusters, as shown in Fig.\ \ref{FiniteT_CMF} (d), is performed including
the clusters up to $2\times 4$ (where the single-site MF value is excluded). 
Additional $(t/U)_c$ data points obtained from a $3\times 3$ cluster 
lie on the respective fitted (dashed) lines which confirms the accuracy
of the extrapolation. Therefore, we will confine the extrapolation procedure
to clusters of size up to $2\times 4$ in the following.

Very close to the QCP, the behavior of $T_c/U$ is difficult to capture, 
however, from our $\lambda\to 1$ extrapolated data we 
observe that $T_c/U$ vanishes linearly with $\delta$ as the
MI-SF transition point is approached, see Fig.~\ref{FiniteT_PD}. 
Such a linear behavior has also been
demonstrated experimentally near the quantum critical point for a
vacuum-to-superfluid transition \cite{Chin12}. On the MI side, we define the
crossover temperature $T^*/U$ between MI and NF as the $\lambda\to 1$ 
extrapolated particle or hole gap $\Delta_{\rm ph}/U$. A true Mott insulator 
with vanishing compressibility $\kappa$ only exists at $T/U=0$. 
The complete phase diagram is shown in Fig.~\ref{FiniteT_PD}. 
We note in passing that the particle
(hole) gap can also be extracted by measuring the distance of a point inside
the Mott lobe from its upper (lower) boundary at a fixed $t/U$, see the 
zero-temperature phase diagram in Fig.\ \ref{CMF_schematic} (b). 
The results of the CMF theory may be verified experimentally by the
quantum gas microscopy technique \cite{Kuhr10, Greiner09}, since it allows 
the measurement of non-local density correlations \cite{Greiner19}.

\begin{figure}[t]
\centering
\includegraphics[width=\columnwidth]{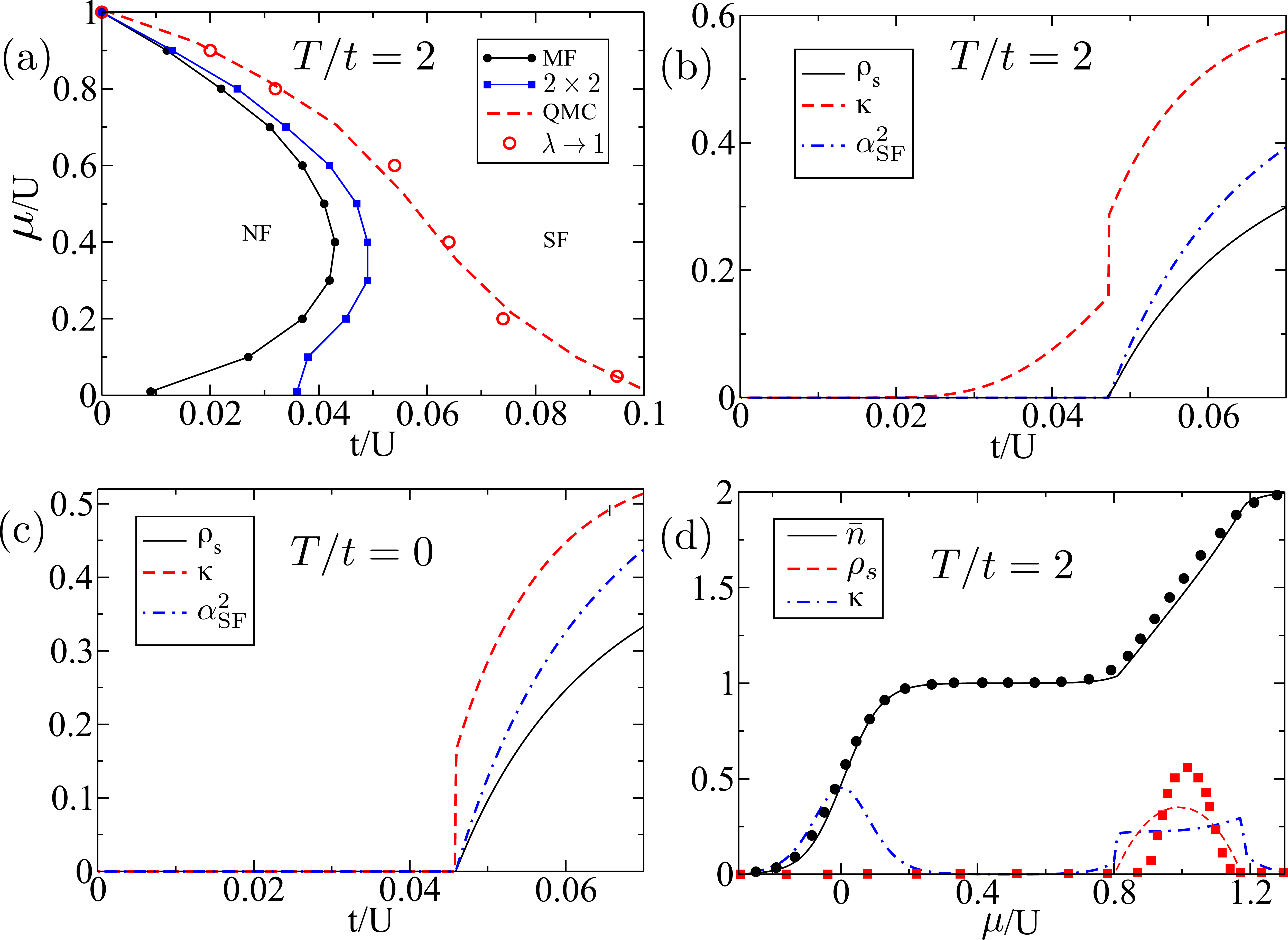}
\caption{(Color online) (a) SF-NF phase boundary of the 2D BHM in the
  $t/U-\mu/U$ plane at a finite temperature $T/t=2$ for different cluster
  sizes including MF and $\lambda \rightarrow 1$ extrapolation. (b, c) The
  compressibility $\kappa$, SF density $\rho_s$ and squared condensate amplitude 
  $\alpha_{\rm SF}^2$ vs. $t/U$ for
  $\mu/U = 0.5$ at $T/t = 2$ and $T/t = 0$ respectively. (d) Average density $\bar{n}$, 
  $\kappa$, and $\rho_s$ as a function of $\mu/U$
  at fixed $t/U = 0.025$ and $T/t = 2$. The plateau marks the first Mott lobe.
For (b-d) we have used a typical cluster size $2\times 2$. Open circles in (a) and filled symbols in (d) are QMC data extracted from Ref.~\cite{Trivedi11}.}
\label{mu_t_PD}
\end{figure}

\subsection{Superfluid fraction and sound velocity in SF phase}
\label{sec:SF_fraction}

In an interacting Bose gas, the superfluid transport is characterized by the 
superfluid density rather than the condensate amplitude. It can be computed
by imposing a phase gradient, which is accomplished by substituting the
hopping amplitude $t$ by $t_{ij}e^{i\theta}$ in Eq.~(\ref{Ham}), where $i,\,j$
are nearest neighbor sites in $x$ direction and $\theta$ is the twist per
lattice site \cite{Fisher73}. Within CMF theory and assuming homogeneity of the
system, the energy of the twisted Hamiltonian can be calculated from a
representative cluster $\mathcal{C}$ only, which yields the following
expression for the SF density \cite{Roth03}, 
\begin{equation}
\rho_s = \frac{F(\theta)-F(0)}{N_{\mathcal{C}}t\theta^2}, \quad 
F(\theta) = - T\ln \mathcal{Z}_C(\theta)
\end{equation} 
where $F(\theta)$ and $\mathcal{Z}_C=\sum_{\nu}\exp[-\beta E_{\nu}(\theta)]$ 
are the free energy and the cluster
partition function, respectively, corresponding to the phase-twisted cluster
Hamiltonian $\hat{\mathcal{H}}_{\mathcal{C}}(\theta)$ with 
twist angle $\theta$ as defined above. 
The eigenvalues $E_{\nu}(\theta)$ of $\hat{\mathcal{H}}_{\mathcal{C}}(\theta)$ can be calculated using the unperturbed eigenstates $|\nu\rangle$ of $\hat{\mathcal{H}}_{\mathcal{C}}(0)$ by means of perturbation theory \cite{Roth03},
\begin{equation}
E_{\nu}(\theta) = E_{\nu}(0) + \theta^2 \sum_{\nu^{\prime} \neq \nu} \frac{|\langle \nu| \hat{J}|\nu^{\prime}\rangle|^2}{E_{\nu}(0)-E_{\nu^{\prime}}(0)} - \frac{\theta^2}{2} \langle \nu| \hat{\mathcal{T}}| \nu\rangle
\end{equation}
where $\hat{J}$ and $\hat{\mathcal{T}}$ are the current and kinetic energy
operators, respectively, which are defined within CMF theory as follows.
\begin{eqnarray}
\hat{J} &=& it\bigg[\sum_{\langle i,j\rangle \in \mathcal{C}}
  \hat{a}_i^{\dagger}\hat{a}_j + \sum_{\langle i,j\rangle, ~i \in
    \mathcal{C}, j\notin \mathcal{C}} \hat{a}_i^{\dagger}\langle\hat{a}_j\rangle - \text{h.c.} \bigg] \\
\hat{\mathcal{T}} &=& \ t\bigg[\sum_{\langle i,j\rangle \in \mathcal{C}}
  \hat{a}_i^{\dagger}\hat{a}_j + \sum_{\langle i,j\rangle, ~i \in \mathcal{C},
  j\notin\mathcal{C}} \hat{a}_i^{\dagger}\langle\hat{a}_j\rangle + \text{h.c.} \bigg] 
\end{eqnarray}
We note that within CMF theory in the pure (non-disordered) BHM the SF density $\rho_s$ vanishes
simultaneously with $\alpha_{\rm SF}$ at the transition to the MI or NF phase
[Fig.~\ref{mu_t_PD}(b,c)].

At the BKT transition in the thermodynamic limit $\rho_s$ has a discontinuous jump,
while the compressibility $\kappa$ exhibits a kink \cite{Duchon13, Rigol12}.
In CMF theory, because vortex fluctuations are not taken into account, $\rho_s$ vanishes
linearly, while compressibility exhibits a jump at the SF-NF boundary even at finite temperature, as shown in Fig.~\ref{mu_t_PD} (b). 
Based on the vanishing of $\rho_s$, or equivalently 
$\alpha_{\rm  SF}$, we obtain a phase diagram in the $\mu/U-t/U$ plane at 
a finite temperature, shown in Fig.\ \ref{mu_t_PD} (a). 
Clearly, the CMF method, after performing the
$\lambda \rightarrow 1$ extrapolation, produces the phase boundary in
quantitative agreement with the QMC, see also Fig.\ \ref{FiniteT_PD}. Thus, it appears to give an overall correct description, except for the critical region of the BKT transition.

\begin{figure}[t]
\centering
\includegraphics[width=\columnwidth]{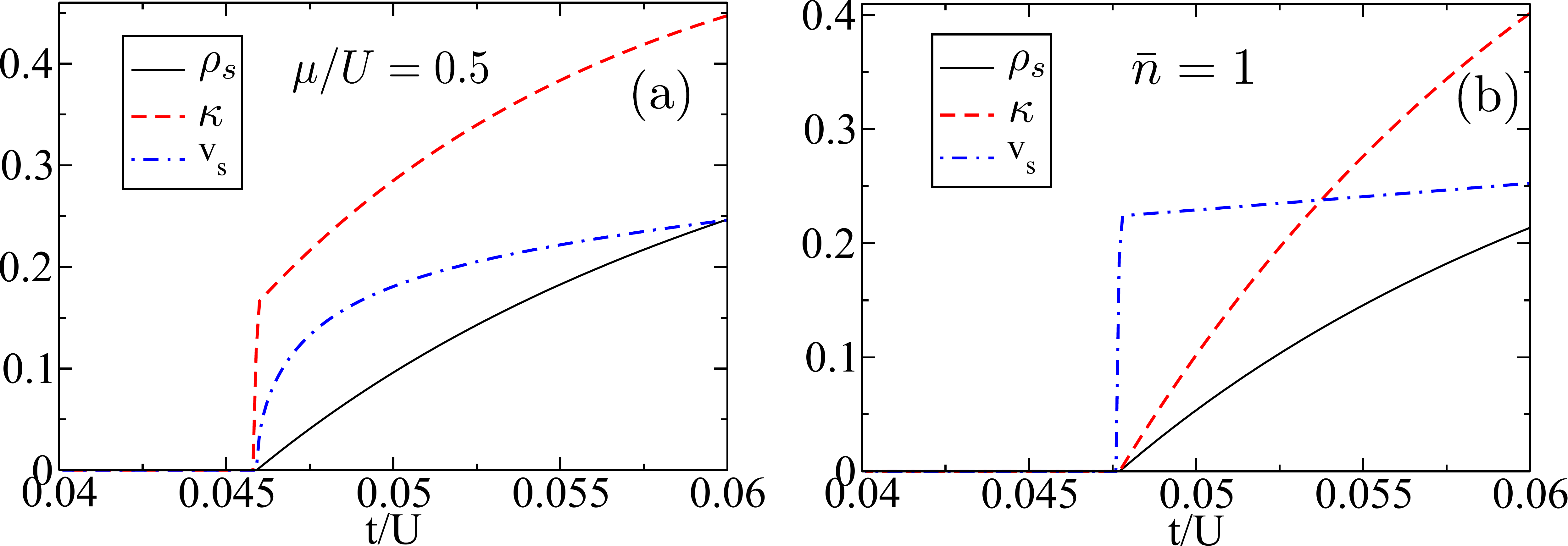}
\caption{(Color online) Compressibility $\kappa$, superfluid density
  $\rho_s$, and sound velocity $v_s$ vs. $t/U$ across (a) a generic
  MI-SF transition for $\mu/U = 0.5$ and (b) across the p-h symmetric Mott tip for a constant density $\bar{n} = 1$. The results are shown for a typical $2\times 2$ cluster.}
\label{chi_fs_vs}
\end{figure}
 
A true incompressible Mott insulator exists only at zero temperature 
where both $\kappa$ and $\rho_s$ vanish [Fig.~\ref{mu_t_PD} (c)].
In Fig.~\ref{mu_t_PD} (d) we also plot the average boson density $\bar{n}$,
SF density $\rho_s$ and compressibility $\kappa$ versus $\mu/U$ for a small
hopping $t/U=0.025$ and a typical $2\times 2$ cluster. Since the effect of
correlations is less for small $t/U$, we indeed observe a good quantitative
agreement with QMC even with a finite cluster. However, for larger $t/U$ 
and a $2\times 2$ cluster the agreement is not as good any more.

Another characteristics of a superfluid with short-range interaction 
is the presence of Goldstone mode
which gives rise to sound modes with finite velocity. The sound velocity is related to the SF fraction and compressibility by a relation \cite{Fisher89},
\begin{equation}
v_s = \sqrt{\frac{\rho_s}{m\kappa \bar{n}^2}}\ ,
\label{sound_velocity}
\end{equation}
where $m=1/2t$ is the boson effective mass in the lattice. 
Fig.~\ref{chi_fs_vs} (a) shows $\rho_s$, $\kappa$, and the resulting $v_s$
as a function of $t/U$ at $T=0$ near a generic, non-p-h symmetric 
MI-SF transition. Within the CMF method the sound velocity vanishes 
$v_s\sim \sqrt{|\delta|}$, since $\rho_s$ vanishes linearly, but $\kappa$ remains finite. 

\subsection{Entanglement entropy}
\label{Entanglement}

\begin{figure}[t]
\centering
\includegraphics[width=\columnwidth]{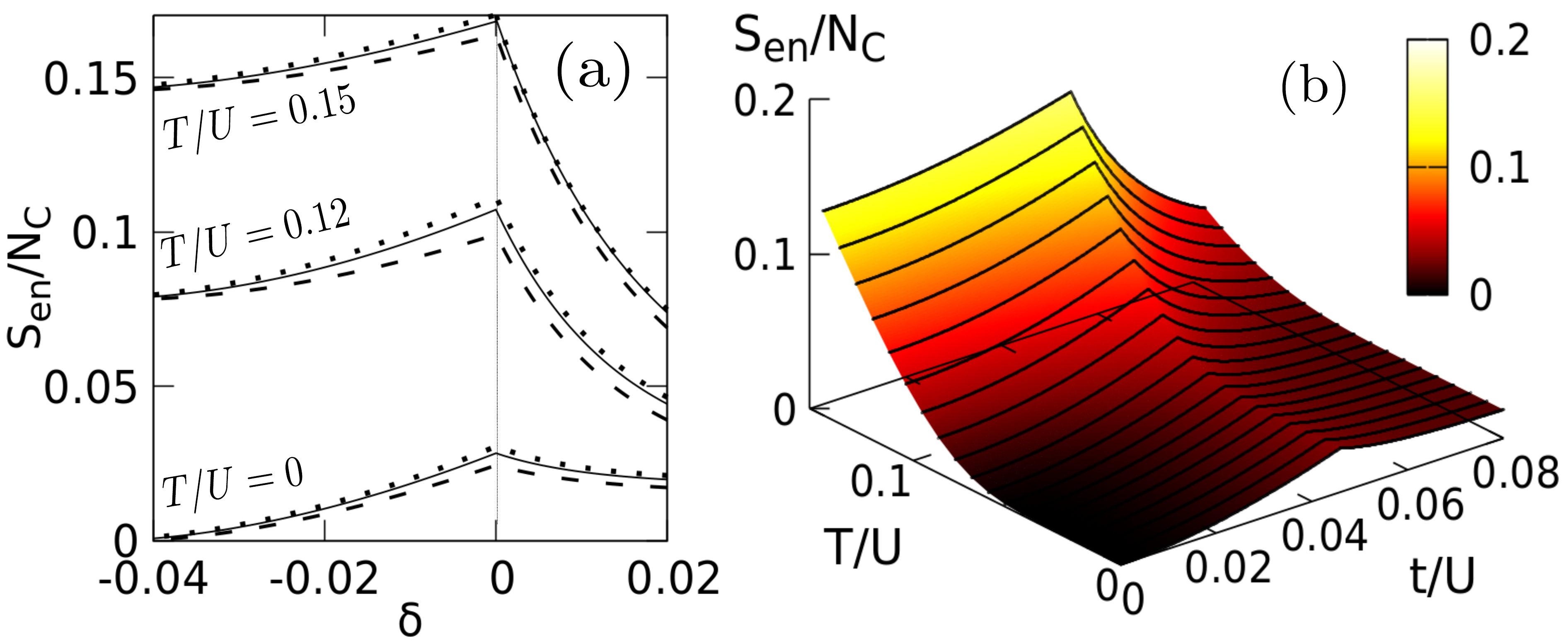}
\caption{(Color online) (a) CMF theory results for the cluster entanglement entropy per site, $S_{\rm en}/N_{\mathcal{C}}$, as a function of $\delta=(t/U)-(t/U)_c$ for three different clusters, $1\times 2$ (dashed line), $2\times 2$ (solid line) and $2\times 3$ (dotted line) at $\mu/U = 0.5$, and for temperatures $T/U$ as shown. (b) Temperature dependence of $S_{\rm en}/N_{\mathcal{C}}$
    for a representative $2\times 2$ cluster across the SF-NF transition.} 
\label{EE}
\end{figure}

In a many-body system, entanglement plays a crucial role in many contexts,
especially near phase transitions where fluctuations are enhanced. In order
to further characterize the SF-NF transition,  
we compute the bipartite entanglement entropy of a cluster, termed, {\it cluster entanglement entropy} (CEE), using CMF theory at finite temperature. 
It can be noted that such a calculation necessarily includes
the thermal contribution since the density matrix depends on
temperature \cite{Chen14, Swingle13, Sinha18}.
It is defined in the following way.
The finite cluster is divided into two parts, $A$ and $B$, which allows us
to write down any state vector in the cluster, e.g., an eigenstate $|\nu\rangle$ of $\hat{\mathcal{H}}_{\mathcal{C}}$ as \cite{Schollwoeck11},
\begin{equation}
|\nu\rangle = \sum_{\alpha,\beta} C^{\alpha\beta}_{\nu} |\alpha\rangle \otimes |\beta\rangle
\label{prod_expansion}
\end{equation} 
where $\{|\alpha\rangle\} \left(\{|\beta\rangle\}\right)$ is a set of basis 
states of the $A$ ($B$) subspaces, and the matrix of expansion coefficients $C_{\nu}$ has dimension $\sqrt{N}$, $N$ being the dimension of
$\hat{\mathcal{H}}_{\mathcal{C}}$. To obtain the CEE 
at temperature $T$, we calculate the reduced density 
matrix of, say, subsystem $A$, 
\begin{equation}
\hat{\rho}^{\rm A} = {\rm Tr}_B \left(\hat{\rho}_{\mathcal{C}}\right), \quad
\hat{\rho}_{\mathcal{C}} = \frac{1}{\mathcal{Z_{\mathcal{C}}}} \sum_{\nu}
e^{-E_{\nu}/T}\,|\nu\rangle \langle \nu |  \ ,
\label{reduced_rho}
\end{equation}
where ${\rm Tr_B}$ is the trace over subsystem $B$.
Inserting Eq.~(\ref{prod_expansion}) into Eq.~(\ref{reduced_rho})
we obtain the following expression for $\hat{\rho}^{A}$ and finally for 
the von Neumann CEE $S_{\rm en}$,
\begin{eqnarray}
\left(\hat{\rho}^{A}\right)^{\alpha\alpha'} &=& \sum_{\nu}
\sum_{\beta}\,C_{\nu}^{\alpha\beta} {C_{\nu}^{\beta\alpha'}}^* e^{-E_{\nu}/T} \\ 
S_{\rm en} &=& -{\rm Tr} \left(\hat{\rho}^{A} \ln \hat{\rho}^{A}\right)
\end{eqnarray}
Since the entropy is an extensive quantity,
the CEE per lattice site should
be essentially independent of cluster size and should, therefore, represent the
bipartite entanglement entropy per site (entropy density) of the macroscopic
lattice. 
This is confirmed in Fig.~\ref{EE} (a), where the CEE per lattice site
is shown across the SF-NF transition as function of $\delta = (t/U)-(t/U)_c$
for three different cluster sizes and three different temperatures.
The enhancement at the phase transition is clearly seen, even though
BKT physics is not accounted for.
The small deviations for different cluster sizes may be attributed to
surface effects induced by sites located at the cluster surface.
In Fig.~\ref{EE} (b) we show the
full temperature dependence of the CEE per site as a function of $t/U$
for a $2\times 2$ cluster which may be considered representative for the
entanglement entropy of the macroscopic lattice and thus relevant
for experimental entropy measurements which have become 
possible using quantum gas microscopy \cite{Greiner15, Greiner19}.

\section{Finite temperature behavior at particle-hole symmetry}
\label{ph_symmetry}

We now focus on the line in
parameter space where p-h symmetry is valid. Within the MI phase,
at p-h symmetry, the particle gap and the hole gap are degenerate, so that
particle and hole fluctuations cancel each other. Therefore, at the 
p-h symmetric point the MI-SF transition is not driven by density
fluctuations, and a different universality class than for a generic
MI-SF transition may be expected \cite{Svistunov08}. 
The p-h symmetric line is characterized by the particle number
per site being $\bar{n} = 1$ while $t/U$ is varied. 
We realize this condition at any finite temperature $T>0$ by appropriately 
tuning the chemical potential $\mu/U$ in the grand-canonical ensemble. 
In the Mott phase at $T=0$, where $\bar{n}=1$
within a finite area (the Mott lobe) in the $t/U - \mu/U$ phase diagram,
the p-h symmetric line is obtained in the limit $T\to 0$.
It thus starts in the $\bar{n}=1$ Mott lobe 
at $t/U=0$ (atomic limit) with $\mu/U=0.5$ and intersects at the tip of 
the Mott lobe. 

\begin{figure}[t]
\centering
\includegraphics[width=\columnwidth]{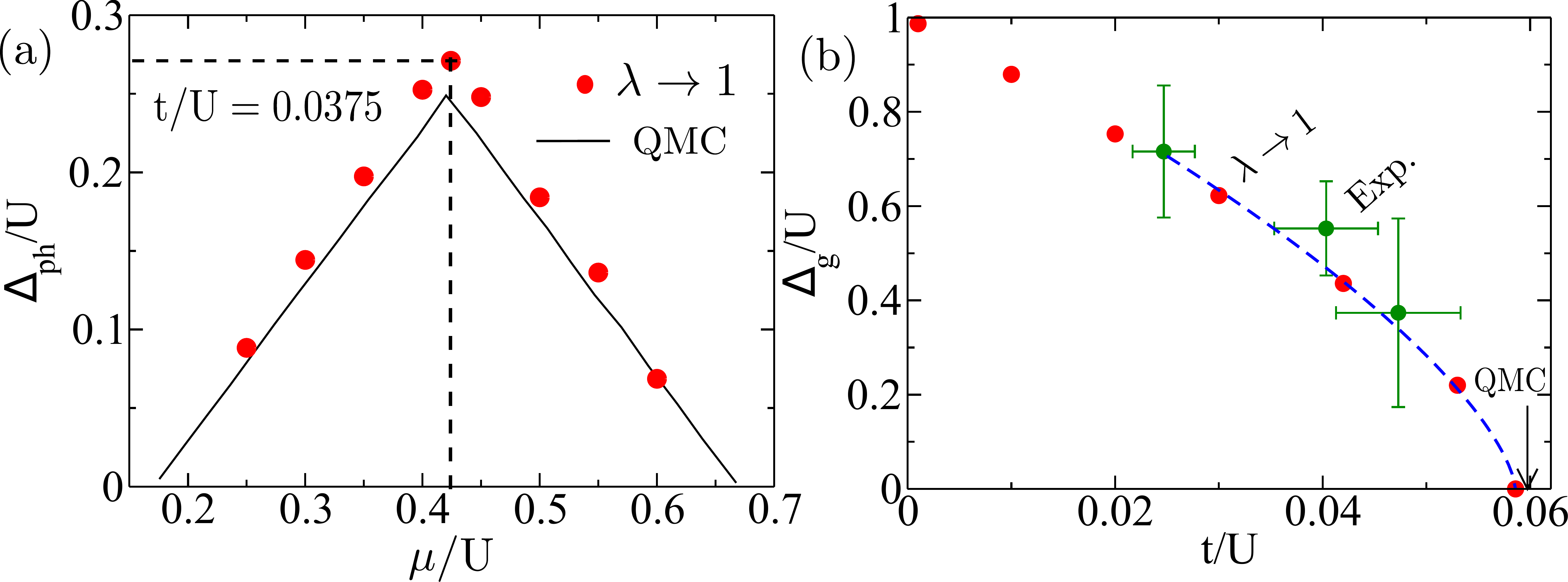}
\caption{(Color online) Particle and hole excitations at the tip of the
  $\bar{n}=1$ Mott lobe. (a) Particle or hole gap $\Delta_{\rm ph}/U$ 
  as a function of $\mu/U$ for $t/U=0.0375$ in comparison to QMC results
  \cite{Svistunov08, Trivedi12}. The dashed lines mark the p-h symmetric 
  value of degenerate $\Delta_{\rm ph}/U$. 
  (b) Mott gap $\Delta_g/U$ as a function of
  $t/U$. Red dots: CMF results in the thermodynamic limit $\lambda\to 1$;
  green dots with error bars: experimental results from Ref.~\cite{Bloch12};
  arrowhead: QMC result from Ref. \cite{Svistunov08}. The 
  $\lambda\to 1$ CMF prediction for the transition point, $(t/U)_c \approx
  0.0578$ is in good agreement with the QMC result of $(t/U)_c \approx
  0.0597$ marked by `$\downarrow$'. The dashed line is a fit of
  $A\delta^{\nu}$ to the CMF data with $\nu=0.6715$ (see text for details).} 
\label{QMC_Exp_extrapolate}
\end{figure}

\subsection{Degenerate particle and hole excitations in the Mott phase}
\label{exp_QMC}

First, we discuss the behavior along the p-h symmetric line within the 
Mott lobe. We extract the gap of particle and hole excitations from the 
temperature dependence of the compressibility as in 
Sec.~\ref{QCP_generic} (c.f. Fig.~\ref{FiniteT_MF}). This defines the 
crossover temperature $T^*=\Delta_{\rm ph}$ between the MI and NF phases. 
In Fig.~\ref{QMC_Exp_extrapolate} (a) we show the particle or hole 
gap $\Delta_{\rm ph}/U$, extrapolated to the thermodynamic limit 
($\lambda\to 1$) as a function of chemical potential
$\mu/U$ for $\bar{n}=1$ at fixed hopping $t/U$ and temperature $T=0$ . 
The results are in good agreement with QMC data. $\Delta_{\rm ph}(\mu)$ reaches its 
maximum value at the p-h symmetric point. 
In experiments, the particle-hole excitation gap $\Delta_g$ is directly
accessible \cite{Bloch12}, while the energy for adding or removing a 
particle, $\Delta_{\rm ph}$, is not. Since $\Delta_g$ is the sum of the 
energies for subtracting and adding a particle, on the p-h symmetric line 
it is just $\Delta_g/U = 2\Delta_{\rm ph}/U$. 
In Fig.~\ref{QMC_Exp_extrapolate} (b) 
we plot $\Delta_g/U$ as a function of $t/U$ along the p-h symmetric line 
($\bar{n}=1$), compared with experimental data as indicated. 
It should be noted that, unlike at a generic MI-SF transition away from
particle-hole symmetry, at the tip of the Mott lobe $\Delta_g$, does
not to vanish linearly. Instead, since in this case the dynamic critical 
exponent is $z=1$, $\Delta_g/U$ must vanish $\sim \delta^{\nu}$ 
[c.f. Eq.~\ref{Delta_critical}], where the correlation-length exponent has been
determined in Ref.\ \cite{Svistunov08} by QMC for the 2D BHM as $\nu = 0.6715$. 
We, thus, fit the Mott gap with a function
$A\delta^{\nu}$, where $A$ is a fit parameter and $\nu = 0.6715$. 
The fitted line is also plotted in Fig.\ \ref{QMC_Exp_extrapolate} (b). 

\begin{figure}[t]
\centering
\includegraphics[width=0.85\columnwidth]{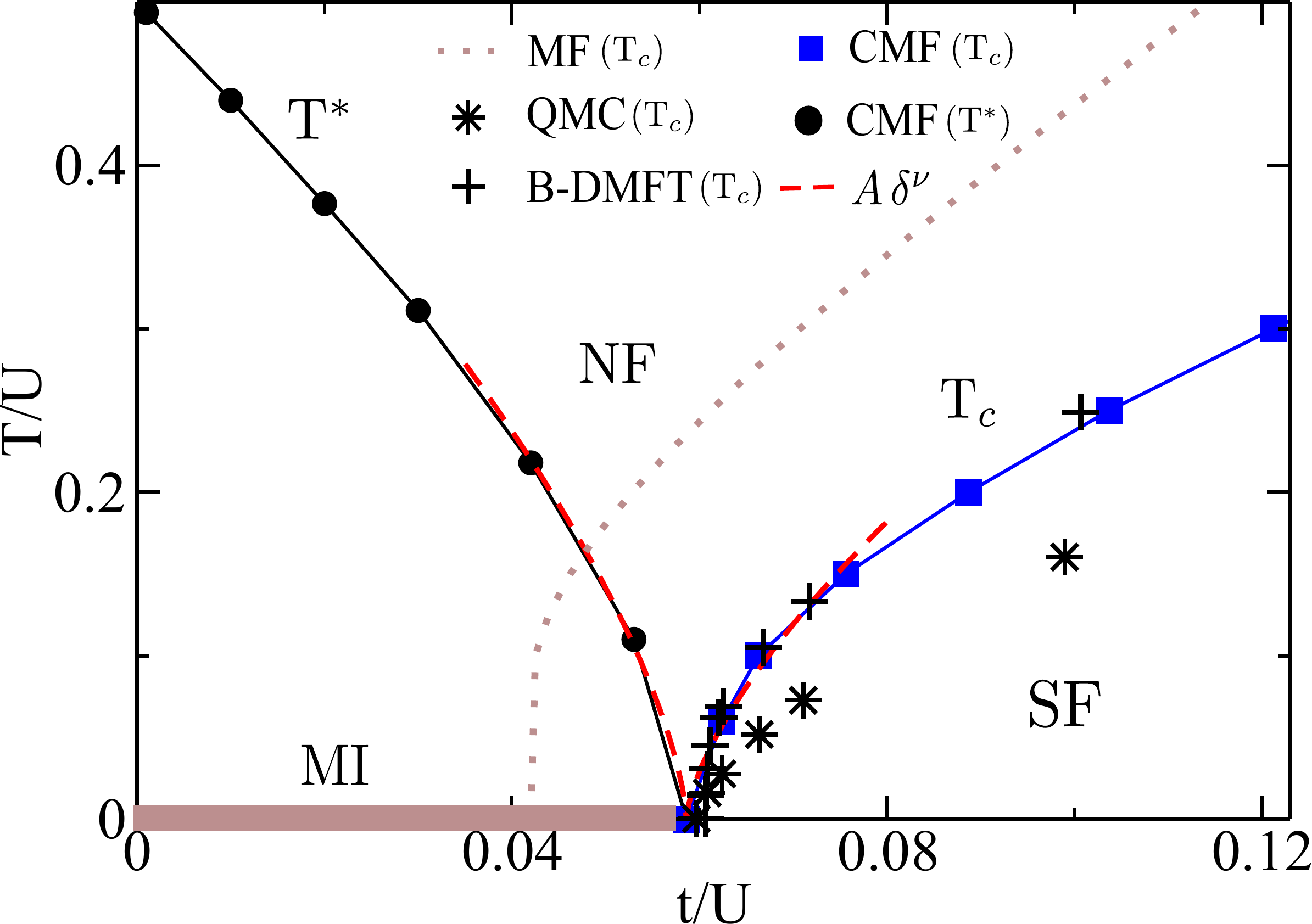}
\caption{(Color online) Finite-temperature phase diagram of the 2D BHM for 
  fixed average density $\bar{n} = 1$. The MI phase at $T/U = 0$ is marked by
  a bold line on the $t/U$ axis. The CMF results for the MI-NF 
  crossover temperature $T^*=\Delta_{ph}$ and the SF-NF transition temperature 
  $T_c$ are shown as black circles and blue squares, respectively. 
  QMC data ($*$) from Ref.~\cite{Svistunov08} and B-DMFT results
  from Ref.~\cite{BDMFT11} are shown for comparison. The red, dashed 
  lines are fits of the function $A\delta^{\nu}$ to our CMF data on the 
  MI as well as the SF side of the transition. The grey, dotted curve represents
  the single-site MF result.}
\label{n1_Tc_fig}
\end{figure}

\subsection{SF-NF transition and phase diagram for $\bar{n}=1$}
\label{n1_Tc}

To complete the phase diagram we now study the SF-NF transition with 
increasing temperature, keeping the average 
density fixed at $\bar{n}=1$. The phase diagram is shown in
Fig.\ \ref{n1_Tc_fig}. The right part shows the critical temperature $T_c/U$ 
line where the condensate amplitude vanishes in the infinite-cluster 
limit $\lambda\to 1$. The crossover line between MI and NF on
the left is obtained as the particle or hole gap $\Delta_{\rm ph}$ of the 
MI at their degeneracy point as discussed in Sec.\ \ref{exp_QMC}, c.f. 
Fig.~\ref{QMC_Exp_extrapolate} (b). 
Fig.~\ref{n1_Tc_fig} also shows QMC \cite{Svistunov08} 
and B-DMFT results \cite{BDMFT11} for comparison. The striking improvement
of CMF over MF theory is clearly seen, particularly, the CMF results are 
consistent with the vanishing of $T^*/U$ as well as $T_c/U$ with an 
exponent of $0.6715$ \cite{Svistunov08} at the QCP, shown as the red, dashed 
lines.  

For the p-h symmetric MI-SF transition (tip of the Mott lobe) both the
compressibility $\kappa$ and the superfluid fraction $\rho_s$ vanish 
at the QCP linearly with $\delta$ as shown using a typical $2 \times 2$ cluster [see Fig.~\ref{chi_fs_vs} (b)].
It follows from  Eq.~(\ref{sound_velocity}) that the sound velocity of the 
superfluid remains finite, in contrast to a generic transition away from 
p-h symmetry as discussed in Sec.~\ref{sec:SF_fraction}, 
c.f. Fig.~\ref{chi_fs_vs} (a).

\section{Summary and conclusion}
\label{summary}

Remarkably, the contrasting behavior of $T_c/U$ between the two universality classes are well captured within CMF theory after performing the cluster size extrapolation to $\lambda \rightarrow 1$.  

In summary, we have computed the finite temperature phase diagram of the
two-dimensional Bose Hubbard model by means of cluster mean-field (CMF) theory
both for a constant chemical potential and for the p-h symmetric case of a
constant density $\bar{n} = 1$. To characterize the different phases, we
computed the condensate amplitude $\alpha_{\rm SF}$ and superfluid density
$\rho_s$ using the CMF theory where both the quantities are finite only 
in the superfluid phase, and the compressibility $\kappa$ which is non-zero 
both in the superfluid and the normal fluid, but vanishes in the Mott insulator. 
While the vanishing of both $\alpha_{\rm SF}$ and $\rho_s$ determines the critical temperature $T_c/U$ for
the superfluid-to-normal fluid transition, the particle or hole gap 
$\Delta_{\rm ph}$ calculated from compressibility defines the crossover temperature
$T^*/U$ between Mott insulator and normal fluid. We further calculated the cluster
entanglement entropy which is enhanced near the critical region and exhibits a
peak at the SF-NF boundary, thereby providing an alternate way to identify the
superfluid transition. We further discussed the behavior of the critical
temperature $T_c/U$ and the crossover temperature $T^*/U$ near the quantum
critical point, both at a generic MI-SF transition, i.e. for a fixed chemical
potential, as well as for a particle-hole (p-h) symmetric transition which
corresponds to the tip of the Mott lobe.

By increasing the cluster size, we analyzed the effect of correlation in
a systematic way. Capturing the variation of the particle/hole excitation gap 
with hopping $t/U$ is a direct consequence of this. We showed how the phase
boundaries got improved with respect to the single-site mean field estimation. 
In the thermodynamic limit, achieved by performing a cluster-size-extrapolation, 
our results quantitatively agree with quantum Monte Carlo (QMC) data and also with experiment. 

Although the critical region of BKT transition is not captured 
by the approach, nevertheless, 
it is remarkable how accurately the CMF method with infinite-cluster 
extrapolation reproduces the critical hopping $(t/U)_c$ and also captures the 
behavior of the superfluid critical temperature $T_c$ and the normal-fluid 
crossover temperature $T^*$ near QCP for both a 
generic and the p-h symmetric transition. This is an
indication that much of the non-local correlations that determine the 
possible critical behavior are incorporated in the finite-size
clusters considered here, and that dynamical fluctuations are of
minor importance here. 

The results of the cluster mean field theory can be tested
experimentally by quantum gas microscopy \cite{Kuhr10, Greiner09}
as discussed in Sec.~\ref{nonlocal_corr} and in Sec.~\ref{Entanglement}. 
Our study further opens up possibilities for future applications of the CMF 
method to investigate non-equilibrium situations like the quench dynamics 
across the quantum phase transition, to analyze the phases of 
bosons in more complicated situations like the Bose glass problem in 
geometrically frustrated or disordered lattices at zero or finite 
temperature.

\appendix

\section{Effect of truncation on Boson number states}
\label{AppenA}

Since the Fock space dimension of bosons is, in principle, infinite, which is, however, 
truncated in any numerical calculations, therefore, it is important to justify such
truncation within the regime of interest. 
\begin{figure}[ht]
\centering
\includegraphics[width=\columnwidth]{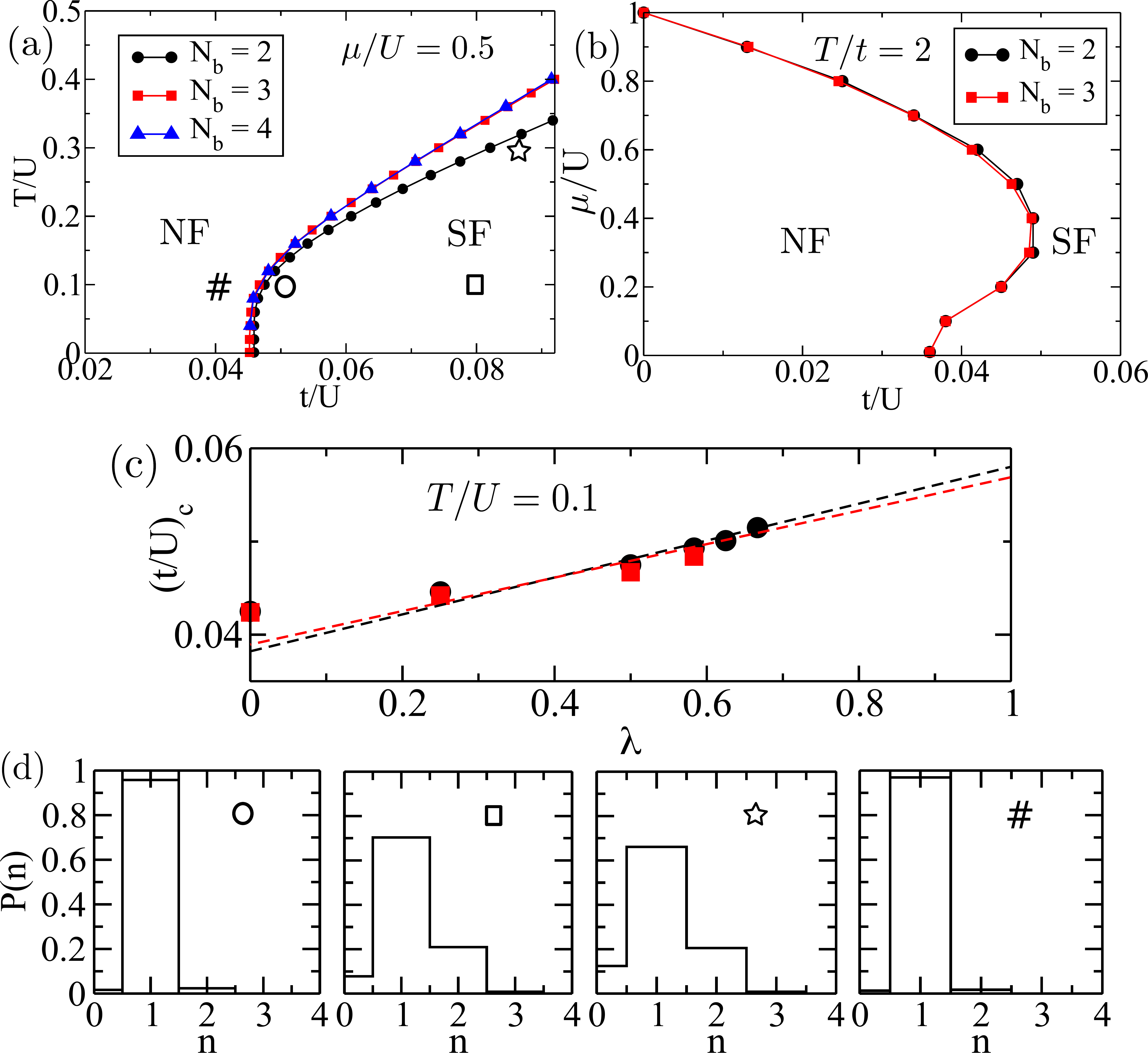}
\caption{(Color online) (a) Phase diagram in $T/U$ vs $t/U$ plane is shown for a generic transition at $\mu/U = 0.5$ using different occupation number truncations $N_b$. In (c) $(t/U)_c$ for different clusters with $N_b = 2$ and $3$ are plotted along with the $\lambda \rightarrow 1$ extrapolation for each of the cases at $T/U = 0.1$. (b) Phase diagram in $\mu/U-t/U$ plane at a finite temperature $T/t = 2$ is shown for number truncations at $N_b = 2, ~3$. (d) Histogram of $P(n)$-distribution over occupation number $n$ in different parts of the phase diagram in (a), as marked by different symbols. The calculations in (a-d), except for (c), are performed using a typical $2\times 2$ cluster.}
\label{n_truncation}
\end{figure}
Since our concern is the accuracy of the 
estimated phase boundary, we plot, in Fig.\ \ref{n_truncation}a, SF--NF phase boundaries 
in $T/U-t/U$ plane for different truncation in the occupation number at $N_b=2,~3$, 
and $4$ using a typical $2\times 2$ cluster. 
Clearly, at low temperatures close to the QCP, which is
the focus of the present study, the curves do overlap for all three cases including
the higher truncation, which is not the case for higher temperature, as expected.
It can be noted that the fitting of $T_c/U$ curve near QCP, as shown in 
Fig.\ \ref{FiniteT_PD} and \ref{n1_Tc_fig}, are done up to temperature $T/U \sim 0.1$, 
and at this maximum point we have overlayed the critical hopping
$(t/U)_c$ for SF--NF transition using $N_b = 3$ on top of the
data obtained using $N_b = 2$ occupation number truncation. For both the cases
the $\lambda \rightarrow 1$ extrapolations are shown. The truncation error on the SF--NF boundary in the extrapolated $(t/U)_c$ turns out to be 
$\sim 2\%$ at $T/U = 0.1$, which is even smaller compared to the standard error
in linear fitting. Therefore, our results near QCP, specifically, the identification of two
different universality classes emerging at and away from the tip of the Mott lobe
are not affected due to number truncation. 
In Fig.\ \ref{n_truncation}b we have further shown that the SF--NF boundaries 
in the $t/U$ vs $\mu/U$ plane for $T/t = 2$ (see blue squares in Fig.\ \ref{mu_t_PD}a 
in the main text) also does not change with increasing the truncation to $N_b=3$.
Further increase in temperature $T/U \sim 0.2$, the phase boundary deviates ($\sim
5\%$ for a $2\times 2$ cluster), as expected, however, that does not concern the critical behavior which is the main purpose of this work.

To further illustrate the effect of truncation in different parts of the phase diagram
we plot the distribution $P(n)$ over the occupation number basis denoted by $n$ in
Fig.\ \ref{n_truncation}d. At low temperature $T/U \sim 0.1$ near the SF--NF boundary (marked by `o' in SF and `\#' in NF) $P(2) \sim 2\%$, which justifies the truncation at $N_b = 2$
in the regime of our interest. However, away from the boundary or at higher temperature 
$P(2) \sim 22\%$, and therefore, a truncation at $N_b = 2$ is no longer sufficient. More
precisely, at higher temperature $T/U \sim 0.3$ the phase boundary, as depicted in Fig.\ \ref{n_truncation}a, is shifted by $\sim 10\%$ in $t/U$ direction. Thereby, the extrapolated values at higher temperatures are also expected to deviate.    

\medskip

\section*{Acknowledgements} 
 
We thank Subhasis Sinha,
Axel Pelster and Anna Posazhennikova for useful discussions. This work was funded by the Deutsche Forschungsgemeinschaft (DFG, German
Research Foundation) under Germany's Excellence Strategy – Cluster of
Excellence Matter and Light for Quantum Computing (ML4Q) EXC 2004/1 --
390534769, and through the DFG Collaborative Research Center CRC 185 OSCAR --
277625399. S.R. acknowledges a scholarship of the Alexander von
Humboldt Foundation, Germany.

\end{sloppypar}

\end{document}